\documentclass[final]{svjour3}

\usepackage{graphicx}
\usepackage{rotating}
\usepackage{amssymb}
\usepackage{mathptmx}
\usepackage[numbers]{natbib}
\makeatletter
\journalname{Journal of Low Temperature Physics}

\usepackage{color}

\newcommand{\blue}{\textcolor{black}}
\bibpunct{}{}{,}{s}{}{,}

\begin{document}

\newcommand{\hdblarrow}{H\makebox[0.9ex][l]{$\downdownarrows$}-}
\title{\blue{Progress in the development of} frequency domain multiplexing for the X-ray Integral Field Unit
on board the Athena mission}

\author{
H. Akamatsu$^{1}$ \and
L. Gottardi$^{1}$ \and
J. van der Kuur$^{2}$ \and
C.P. de Vries$^{1}$ \and
M.P. Bruijn$^{1}$ \and
J.A. Chervenak$^{3}$ \and
M. Kiviranta$^{4}$ \and
A.J. van den Linden$^{1}$ \and
B.D. Jackson$^{1,2}$ \and
A. Miniussi$^{3}$ \and
K. Ravensberg$^{1}$ \and
K. Sakai$^{3}$ \and
S.J. Smith$^{3}$ \and
N. Wakeham$^{3}$
}

\institute{1: SRON Netherlands Institute for Space Research, the Netherlands \\
2: SRON Groningen Netherlands Institute for Space Research, the Netherlands \\
3: NASA GSFC, Greenbelt Road, Greenbelt, MD 20771, USA	\\
4: VTT, Espoo, Finland	\\
\email{h.akamatsu\_at\_sron.nl}}

\titlerunning{Progress in the development of the FDM readout technology for \textit{Athena} X-IFU}
\authorrunning{H. Akamatsu et al. }

\maketitle

\begin{abstract}
Frequency domain multiplexing (FDM) is the baseline readout system for the X-ray Integral Field Unit (X-IFU) on board the Athena mission. Under the FDM scheme, TESs are coupled to a passive LC filter and biased with alternating current (AC bias) at MHz frequencies. Using high-quality factor LC filters and room temperature electronics developed at SRON and low-noise two-stage SQUID amplifiers provided by VTT, we have recently demonstrated good performance with the FDM readout of Mo/Au TES calorimeters with Au/Bi absorbers. We have achieved a performance  requested for the demonstration model (DM) with the single pixel AC bias ($\Delta E=$1.8 eV) and 9 pixel multiplexing ($\Delta E=$2.6 eV) modes. We have also demonstrated 14-pixel multiplexing with an average energy resolution of 3.3 eV, which is limited by non-fundamental issues related to FDM readout in our lab setup. 


\end{abstract}
\vspace{-0.75cm}

\section{Introduction}
\vspace{-0.25cm}
Athena is 2nd L-class mission in the ESA's cosmic vision program\cite{2019arXiv191204615B}. To answer questions like \textit{How does
the ordinary matter assemble into the large-scale structures that we see today? } and \textit{How do black holes grow and
influence the Universe? }, Athena will employ two main focal plane instruments: X-ray Integral Field Unit (X-IFU\cite{XIFU18}) and Wide Field Imager (WFI\cite{WFI13}). The X-IFU instrument consists of an array of $\sim$3000 Transition Edge Sensor X-ray microcalorimeters (TES calorimeter) with a high  spectral resolution $\Delta E$=2.5 eV up to 7 keV ($E/\Delta E\sim$2800). 
Due to strictly limited available electrical and cooling power in space, the multiplexing technology is one of key technology for X-IFU. 

We are developing the frequency domain multiplexing (FDM) read-out of TES calorimeters for the X-IFU instrument. Under the FDM scheme, each TES is connected with a passive LC filter and biased with alternating current (AC bias) at MHz frequencies.
Each resonator should be separated beyond their detector response to avoid crosstalk between neighboring resonators.
To satisfy the requirement of the X-IFU, a multiplexing factor of 40 pixels/channel in a frequency range from 1 to 5 MHz required. Therefore frequency separation for the X-IFU will be 100 kHz. 


\section{NASA/GSFC TES calorimeter array and experimental setup} \vspace{-.25cm}
\subsection{TES X-ray microcalorimeter array}\label{sec:detector} \vspace{-.25cm}
In this paper, we report spectral resolution and multiplexing results of TES calorimeter arrays from NASA/GSFC group. 
The TES is made of Mo/Au bilayer with a BiAu-mushroom type absorber.  The basic information of TES arrays are summarized in Table~\ref{tab:device}. Detailed characterizations of these devices under AC bias is presented in ~\cite{luciano_LTD18_AC}.

\begin{table}[t]
\begin{center}
\caption{Basic information of NAS/GSFC TES calorimeter and our electrical circuits
\label{tab:device}
}
\begin{tabular}{cccccccccc} \hline
				& 	TES size 					&	$R_{\rm N}$	& $T_{\rm c}$	& Critical inductance 		& Transformer 				&  Effective inductance\\ 
				&	[$\mu \rm m^2$]			&	[m$\rm \Omega$]		&	[mK]			& $@$ 20 $\% \rm ~of~ Rn$ [nH]		&	ratio						& [nH] \\ \hline			
Device A6 		&	100$\times$100 			&	35--35		&	89		&	550	& 1:3.0	& 274	\\
Device A7 		&	120$\times$120 			&	30--40		&	87		&	670	& 1:2.0 	& 565\\
\hline
\end{tabular}
\end{center}
\end{table}%

\subsection{Experimental setup}\label{sec:setup} 
For details of the experimental setup, the readers are recommended to refer to previous works \cite{akamatsu14,akamatsu16}. For the completeness,
we are using cryogen  free dilution unit\cite{LCcooler}, which has a cooling power of 400~$\mu$W at 110 mK. 
A Germanium thermistor (Lake Shore GR50) is implemented in the Cu experimental plate, which has a temperature sensitivity $\alpha~(\equiv\frac{T}{R}\frac{dR}{dT})\sim5$. At 60 mK, the base temperature in this report, the typical temperature stability is about 0.5--0.7 $\mu \rm K_{\rm rms}$ over a day\cite{akamatsu16}.

SRON in-house lithographical LC filters are implemented at the same temperature stage of the detector. 
These LC resonators show high Q-factors as large as 10,000\cite{LCfilter18}.
The inductance was designed to be 2 $\mu$H. By tuning capacitance values, 18 resonators span 1--5 MHz. 0.75~$\rm \Omega$ bias shunt resistance and 1:25 capacitive voltage divider are implemented in the detector stage. 
To match SQUID dynamic range and tune damping inductance\cite{2016SPIE.9905E..5RV}, we are using a superconducting transformer with primary inductance of 48 nH. The typical coupling coefficient is about 0.92--0.94.
The transformation ratio for each detector is also given in table~\ref{tab:device}. 
Room temperature high-$\mu$ metal and superconducting Nb are employed for the magnetic shielding. 
Superconducting Helmholz coils are implemented in the experimental plate to cancel the remnant magnetic field and investigate magnetic field dependence, respectively.
\blue{For the spectral performance evaluation, }
$^{55}$Fe source is mounted on the Nb shield. The count rate is tuned to be $\sim$1 cps. 
For the FDM readout, SQUID is one of the most important component. We are using two stage VTT SQUIDs amp consisting of
6 array front end and 184$\times$4 array amplification SQUIDs. 
The mechanical vibrations are damped by Kevlar wires\cite[][]{luciano_vib19}.
To increase the bandwidth of the FDM, we compensate for the delay between the room temperature and 50 mK by using BaseBand Feed Back (BBFB)\cite{BBFB}. The SQUID signal is demodulated and re-modulated with phase shift by an FPGA board before it is feedback. We are using SRON in-house analog (SQUID controller \& LNA) and digital (demodulation, BBFB, etc) electronics. Data were obtained in 40 M samples/s, decimated to 156 k samples/s, and demodulated around the resonant frequency of each LC filter.

\begin{figure}[t]
\begin{center}
\includegraphics[width=0.485\linewidth, keepaspectratio]{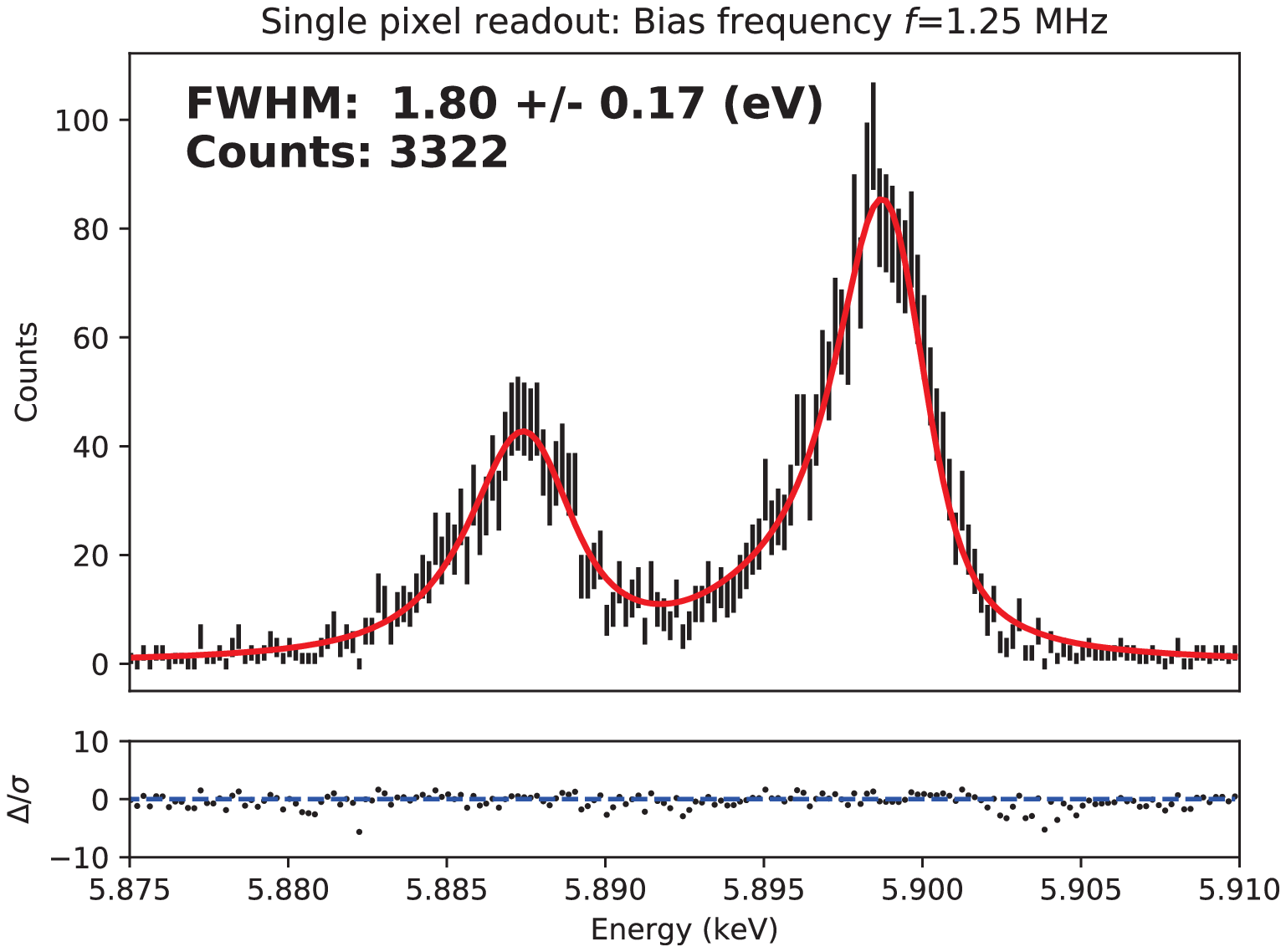}
\includegraphics[width=0.485\linewidth, keepaspectratio]{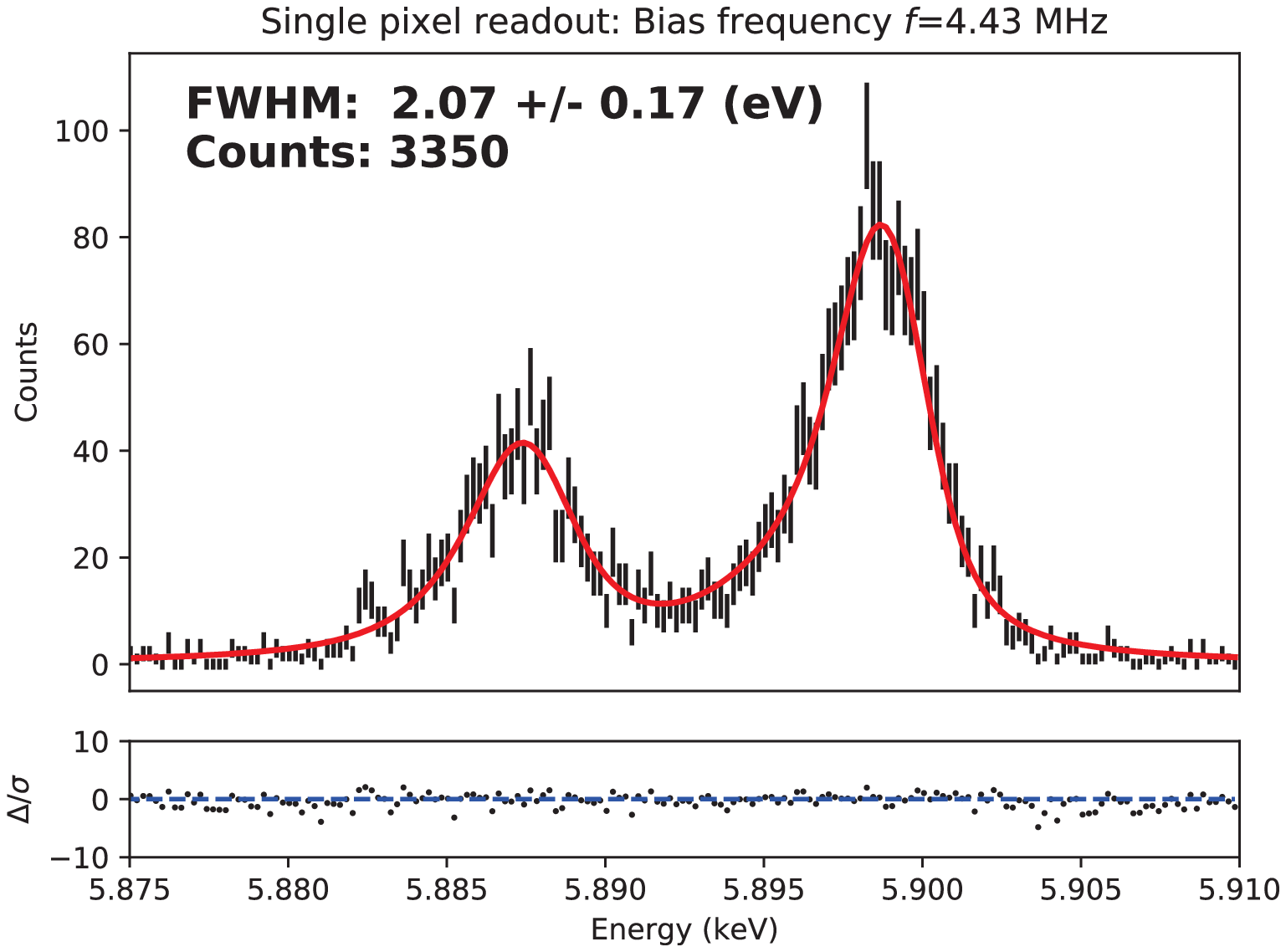}
\caption{\label{fig:single}
Energy spectrum of Mn K$\alpha$ complex from $^{55}$Fe source. Black bar and red solid line indicate measured spectrum and the best-fit model, respectively. 
Bottom panel shows the normalized residual in the unit of  (data-model)/statical error.
Left and right panels show spectrum at AC bias frequency of 1.25 MHz and 4.43 MHz in the single pixel readout mode, respectively.
}
\vspace{-0.75cm}
\end{center}
\end{figure}

\section{Results}\label{sec:results}\vspace{-0.25cm}
\subsection{Performance under single pixel mode \& Multiplexing experiments}\label{sec:single}\vspace{-0.25cm}
To investigate potential limitations, we first start characterizations of the detector performance under AC bias.
For this measurement,  we used Device A6 (100$\mu \rm m^2$ TES). The bias points were set around $R/R_{\rm N}\sim 0.2$. The remnant magnetic field was canceled by the Helmholz coils. 
The long term temperature drift correction was performed by using TES baseline current and pulse hight information.
Using the zero energy (0 keV), Mn-K$_\alpha$ (5.9 keV) and Mn-K$_\beta$ (6.5 keV) information, the energy non-linearity was also corrected. The typical energy non-linearity factor is about 1--2 \%. 
The energy spectra were fitted with the Mn-K$_\alpha$ line model  by minimizing the total C-stat value\cite{cstat} to avoid undesired fitting bias. 

The best performance under AC bias at 1.25 and 4.40 MHz are $\Delta E=1.8\pm0.2$ eV and $2.1\pm0.2$ eV, respectively (Fig. \ref{fig:single}) \blue{although typical range of the performance are $\Delta E\sim 2.0-2.5$ eV}.
At the low bias frequency, for the first time, we have archived DC bias compatible performance under AC bias\cite{smith16, 2018JLTP..193..337M}. 
At higher frequency, there is a slight degradation which can be impacts of AC loss\cite{2018JLTP..193..356S}, AC Josephson effect\cite{2018JLTP..193..209G} and contributions from the room temperature electronics.
Both performances are better than predictions from the integrated noise equivalent power (NEP). 
\blue{Because of the large inductance, the pulse excursion is getting larger. 
This leads that the noise decreases during the pulse. Therefore, better X-ray resolutions than NEP resolution will be achieved.
}
\blue{
Therefore, in this paper, we will mainly focus on X-ray performances.
}

\begin{figure}[t]
\begin{center}
\includegraphics[width=0.485\linewidth, keepaspectratio]{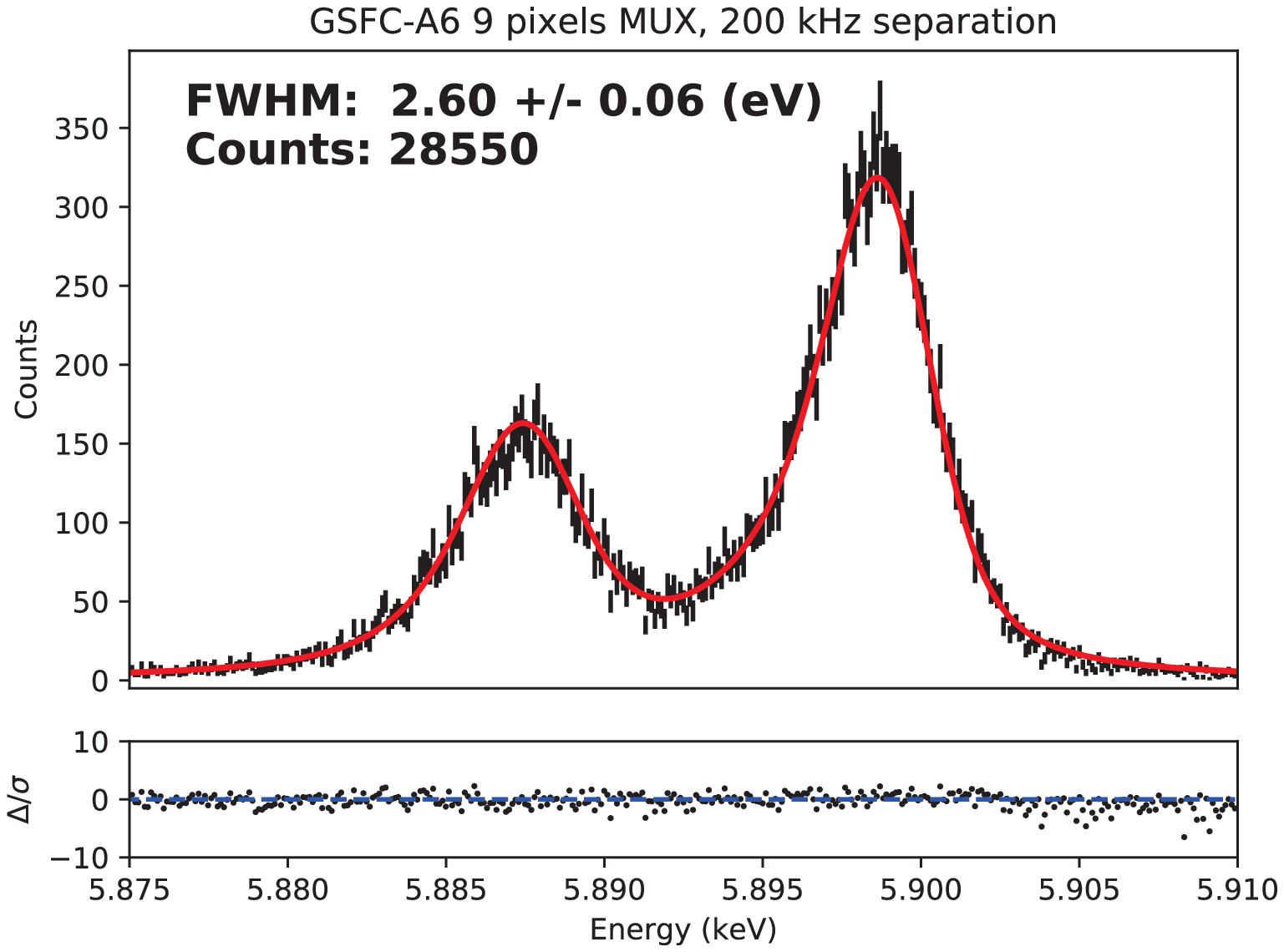}
\includegraphics[width=0.485\linewidth, keepaspectratio]{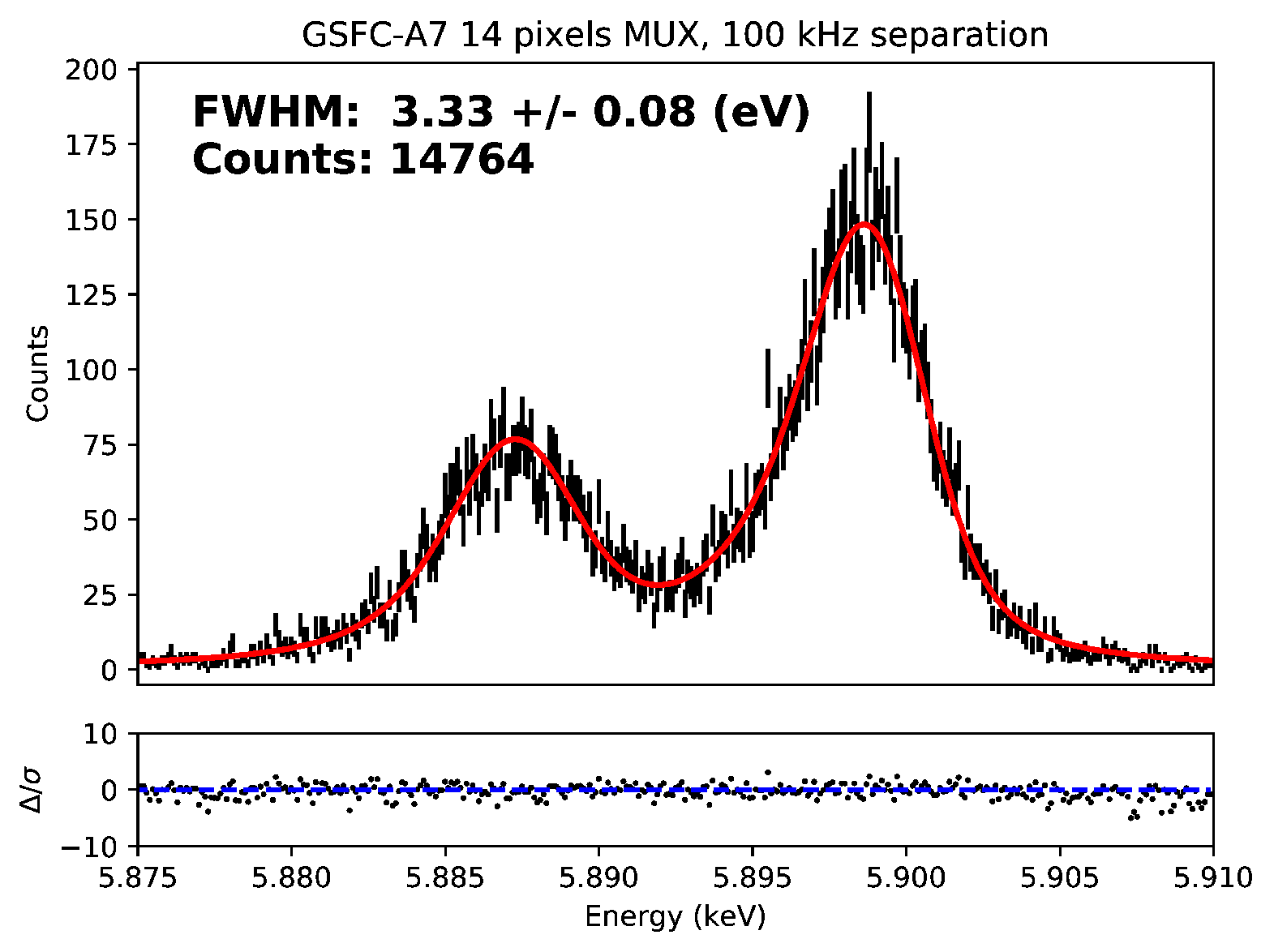}
\includegraphics[width=0.485\linewidth, keepaspectratio]{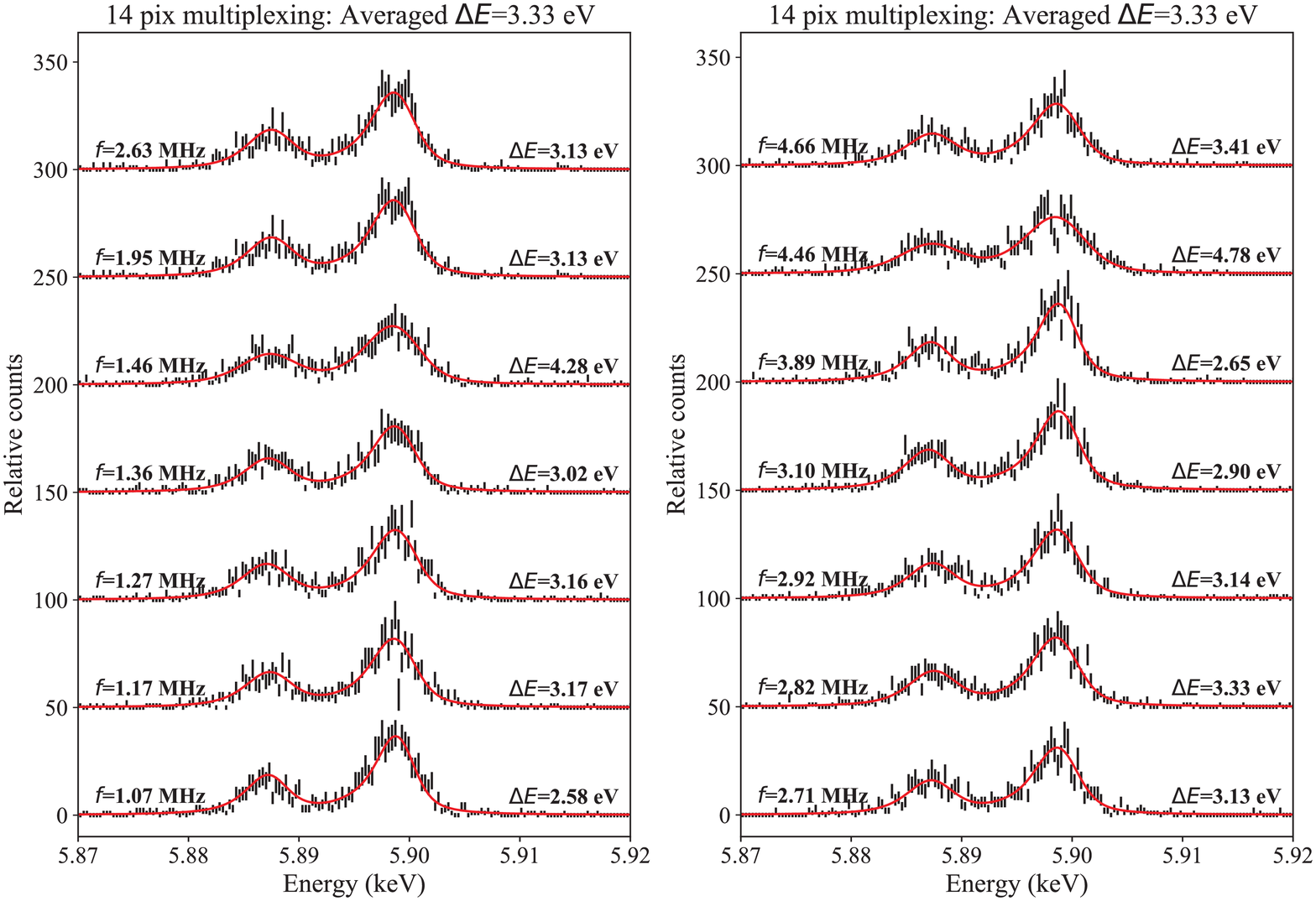}
\includegraphics[width=0.325\linewidth, keepaspectratio]{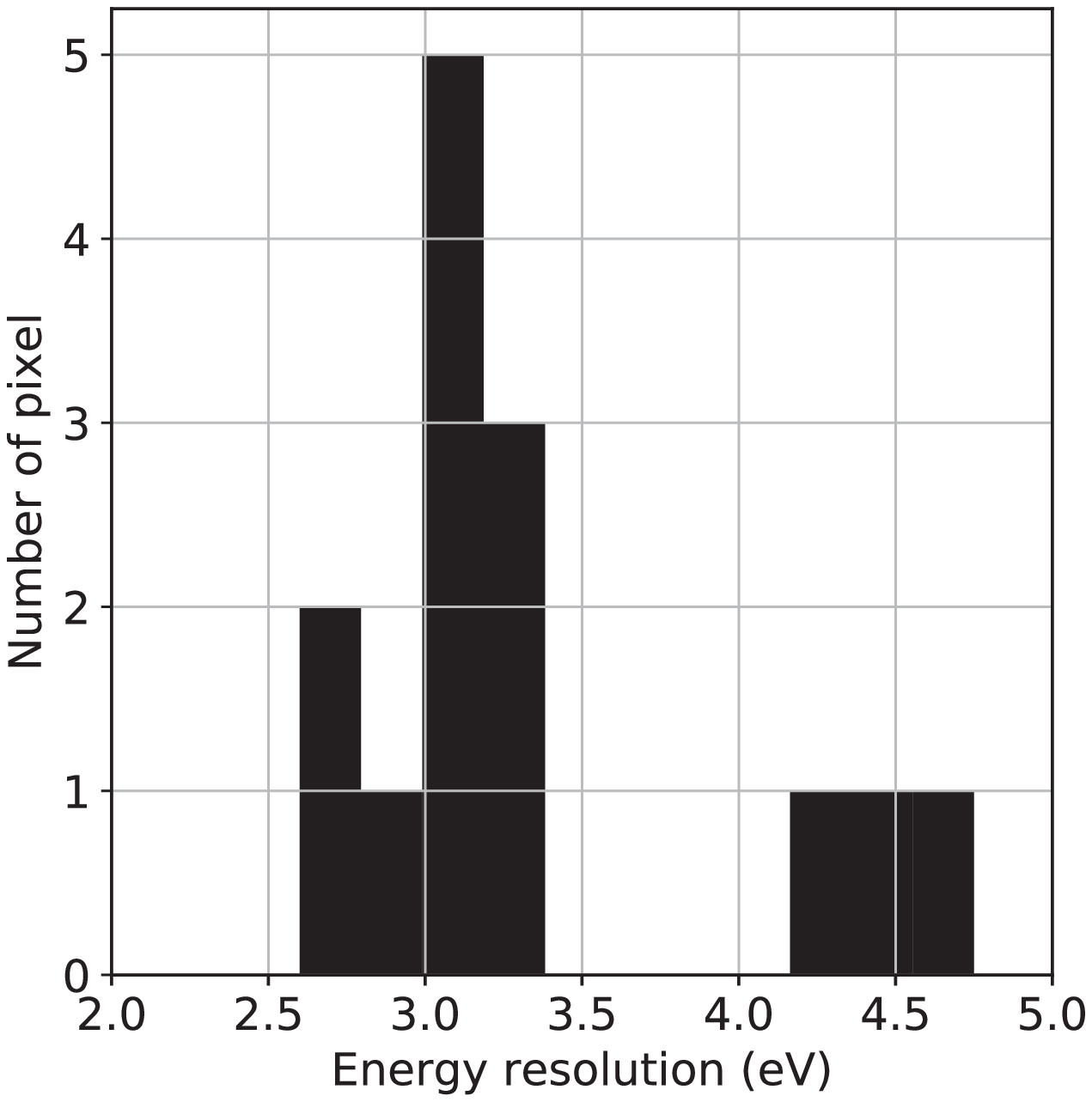}
\caption{\label{fig:MUX}
(Top) Summed spectra of 9 pixel MUX with 200 kHz (Device A6) and 14 pixel MUX with 100 kHz (Device A7), respectively.
(Bottom left) 14 individual spectra in the 14 pixel MUX mode. 
(Bottom right) Histogram of the energy resolution in 0.2 eV bin.
\vspace{-0.5cm}
}
\end{center}
\end{figure}

We perform multiplexing experiments with two different conditions:
1) Device A6 and 9 pixel MUX with 200 kHz separation  and 2) Device A7 and 14 pixel MUX with 100 kHz separation.
The results are shown in Fig.\ref{fig:MUX}. 
The summed spectral resolutions are $\Delta E$=2.6 eV, 3.3 eV for 9 pixel MUX and 14 pixel MUX,  respectively. 
\blue{
In the 200 kHz MUX case,  the summed spectral resolution surpasses the requirement of the demonstration model of X-IFU instrument ($\Delta E < 3 \rm~ eV$) although the frequency separation did not match with the X-IFU requirement (100 kHz separation). 
Therefore, we took an approach to put further increase multiplexing number rather than improving the spectral performance to identify potential bottlenecks with 100 kHz separation and given configurations.
}
In the 100 kHz separation case, the resonator frequencies span across 1.0--4.7 MHz. The summed energy resolution ($\Delta E\sim3.3$ eV)  degrades considerably from 200 kHz separation configuration ($\Delta E\sim$2.6 eV). 
\blue{
The bottom right panel in Fig.2 shows a histogram of energy resolutions. 
}
There are roughly three populations: 1.) around 2.6 eV, 2.) around 3.0 eV and 3.) 4.4--4.6 eV.
We figured out that these populations can be explained by 1.) impacts of intermodulation line noises and 2.) influences from 
neighboring  bias voltages.
\vspace{-0.5cm}

\begin{table}[t]
\begin{center}
\caption{Detector performances  $@$ 6 keV under 3 pixel MUX mode (unit of eV)
\label{tab:3MUX}
}
\begin{tabular}{cccccccccc} \hline
& Single pixel Mode& \blue{3 pix MUX without FSA}& \blue{3 pix MUX with FSA}\\ \hline
Ch0 ($f$=1.07 MHz)	&2.52$\pm$0.18	& 2.78$\pm$0.18	& 2.93$\pm$0.14	\\ 
Ch1 ($f$=1.17 MHz)	&2.20$\pm$0.15	& 5.58$\pm$0.18	& 3.07$\pm$0.16	\\
Ch2 ($f$=1.27 MHz)	&2.39$\pm$0.18	& 2.85$\pm$0.17	& 3.01$\pm$0.16	\\ 
\hline
\vspace{-0.5cm}
\end{tabular}
\end{center}
\end{table}%

\begin{figure}[t]
\begin{center}
\includegraphics[width=0.485\linewidth, keepaspectratio]{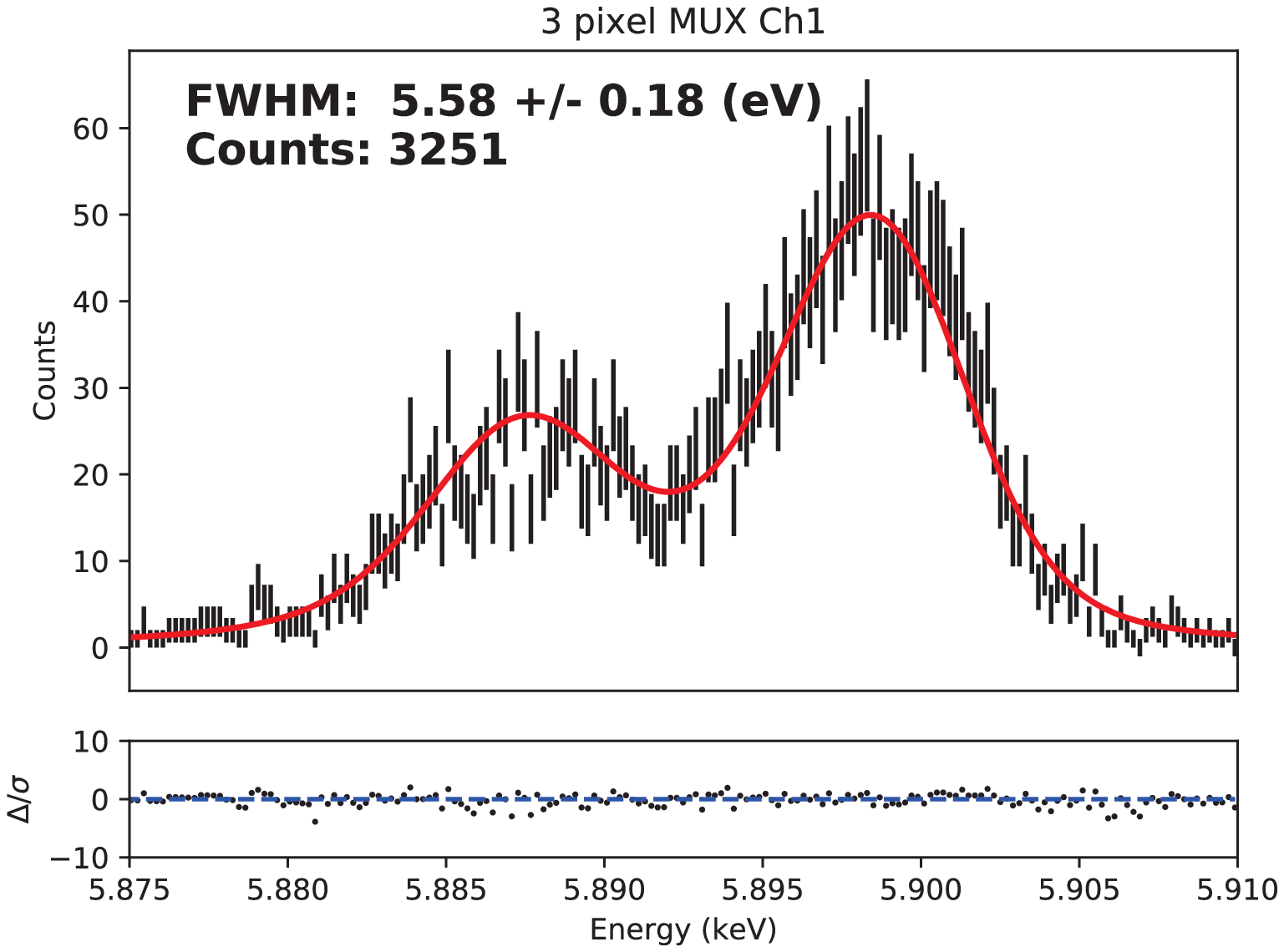}
\includegraphics[width=0.485\linewidth, keepaspectratio]{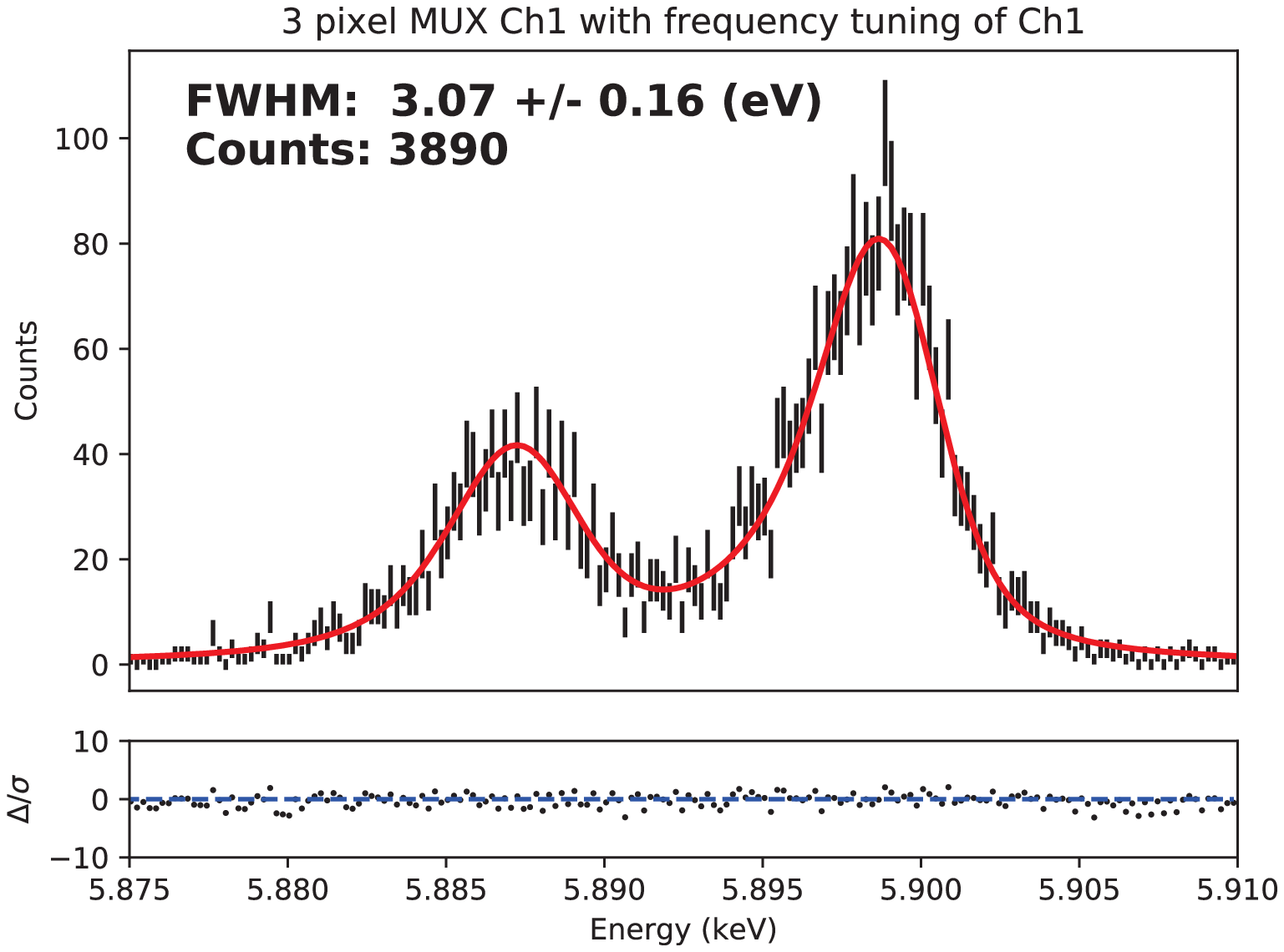}
\includegraphics[width=1\linewidth, keepaspectratio]{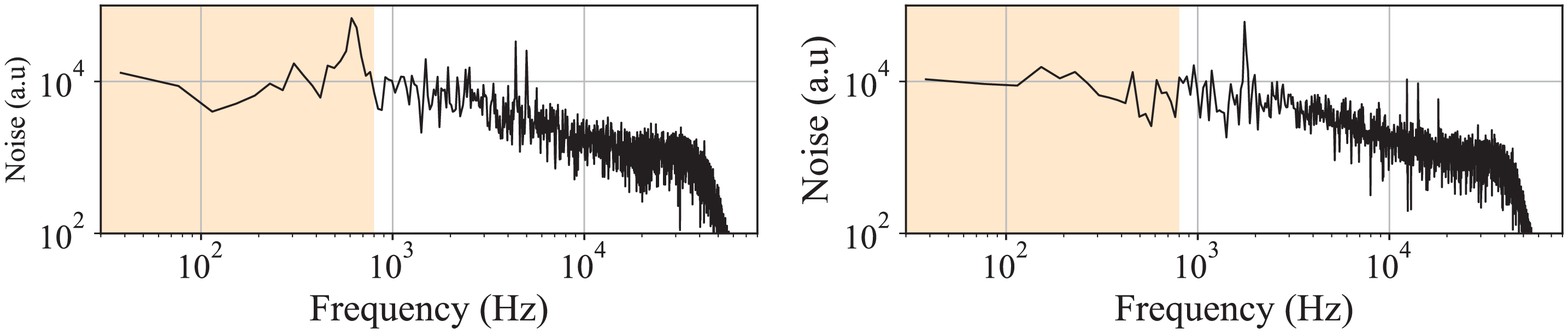}
\caption{ \label{fig:3MUX}
(Top) Under 3 pixel MUX mode, spectra of Ch1 ($f$=1.17 MHz) without (left) and with FSA (right).
(Bottom) Noise spectra of Ch1 without (left) and with FSA (right). The orange shed area indicates 
two times the detector's thermal band.
}
\vspace{-0.75cm}
\end{center}
\end{figure}

\subsection{Impact of the intermodulation line noise}\label{sec:line}\vspace{-0.25cm}
\blue{Any non-linearities will generate spurious line noises via the intermodulation mechanisms.}
Such non-linearity can be caused by the SQUID amplifiers, the DAC's,  and even the passive components on the board.
\blue{For the worse performance pixels ($\Delta E\sim$4.5 eV), we confirmed the presence of line noises within the thermal band.}

For the DAC case, the effect of the non-linearity can be removed, by using carriers in a frequency arrangement where the frequency spacing between all subsequent carriers is the same and where each carrier frequency is an integer number times that spacing. In this way, all distortion products will fall on a carrier and no spurious line noises will be present between the carriers. 

To confirm the impact of the line noises, we performed an additional 3 pixel MUX experiment with 100 kHz separation configuration. The resonant frequencies are $f$=1.07 MHz (Ch0), 1.17 MHz (Ch1) and 1.27 MHz (Ch2).
In the single pixel mode, all three pixels show good performances around 2.2-2.5 eV (Table~\ref{tab:3MUX}). 
Under 3 pixel MUX mode (Fig.\ref{fig:3MUX} top left), despite  the other two pixels show moderate changes, 
the central pixel (Ch1: $f$=1.17 MHz) shows significant degradation from $\Delta E=$2.20 eV to 5.58 eV.
As expected, there is a significant line noise component in the Ch1 noise spectrum (Fig.\ref{fig:3MUX} bottom left). The line noise is located at the frequency, in which the detector is still sensitive (orange shaded area), meaning the detector feels modulation at line noise frequency in their bias voltage. 

The impact of this intermodulation line noise can be avoided by changing bias frequency. However, due to the high-Q LC filter, changing bias frequency generates a different electrical circuit. As shown in the left panel of Fig.\ref{fig:FSA} (gray points), the detector performance has a strong bias frequency dependency.
To mitigate this dependency,  we are developing  the frequency shift algorithm (FSA). The details of FSA are given by the previous works\cite{2018JLTP..193..626V}. Under the FSA scheme, the on-resonance detector performance and response will be kept with different bias frequencies. The red points in Fig.\ref{fig:FSA} show the performance with FSA with different bias frequencies.
Although the detector performance across wide frequency range will be available with FSA, 
it still needs fine tuning to \blue{
obtain same performance at the resonance frequency.
}

Back to the 3 pixel MUX experiment, we applied FSA to Ch1 with 400 Hz shift to kick the intermodulation line noise out from the detector thermal band. 
\blue{As shown in th eright top panel in Fig.\ref{fig:3MUX}}, FSA improves the performance significantly from 5.6 eV to 3.1 eV without any influence on the other two pixels within given statistical errors (Table.\ref{tab:3MUX}). Noise spectrum after frequency shift shows a clear difference in term of the location of the line noise, which shifted toward a higher frequency regime. 
\vspace{-.75cm}

\subsection{Influences from neighboring AC bias voltages}\label{sec:ACB}\vspace{-0.35cm}
As demonstrated in Sect.\ref{sec:line}, the 3 pixel MUX experiment also shows around 3 eV performance after removing the impact of the intermodulation line noise. It is consistent with the highest peak in the bottom right panel of Fig.\ref{fig:MUX}.
We figured out that the degradation to $\sim$3 eV could be the interferences from neighboring AC bias voltages.
The effect was already reported in our earlier works\cite{2003ITAS...13..638V}. 
If the detector electrical bandwidth ($\sim R/L$) is too large, the detector is still sensitive to neighboring AC bias voltage, resulting the modulations in TES current with $\Delta f$ (resonant frequency -  neighboring bias frequency) interval. In this case, the integrated NEP resolution will not be affected. but X-ray resolution will be degradated by the modulations. 

\begin{figure}[t]
\begin{center}
\includegraphics[width=0.485\linewidth, keepaspectratio]{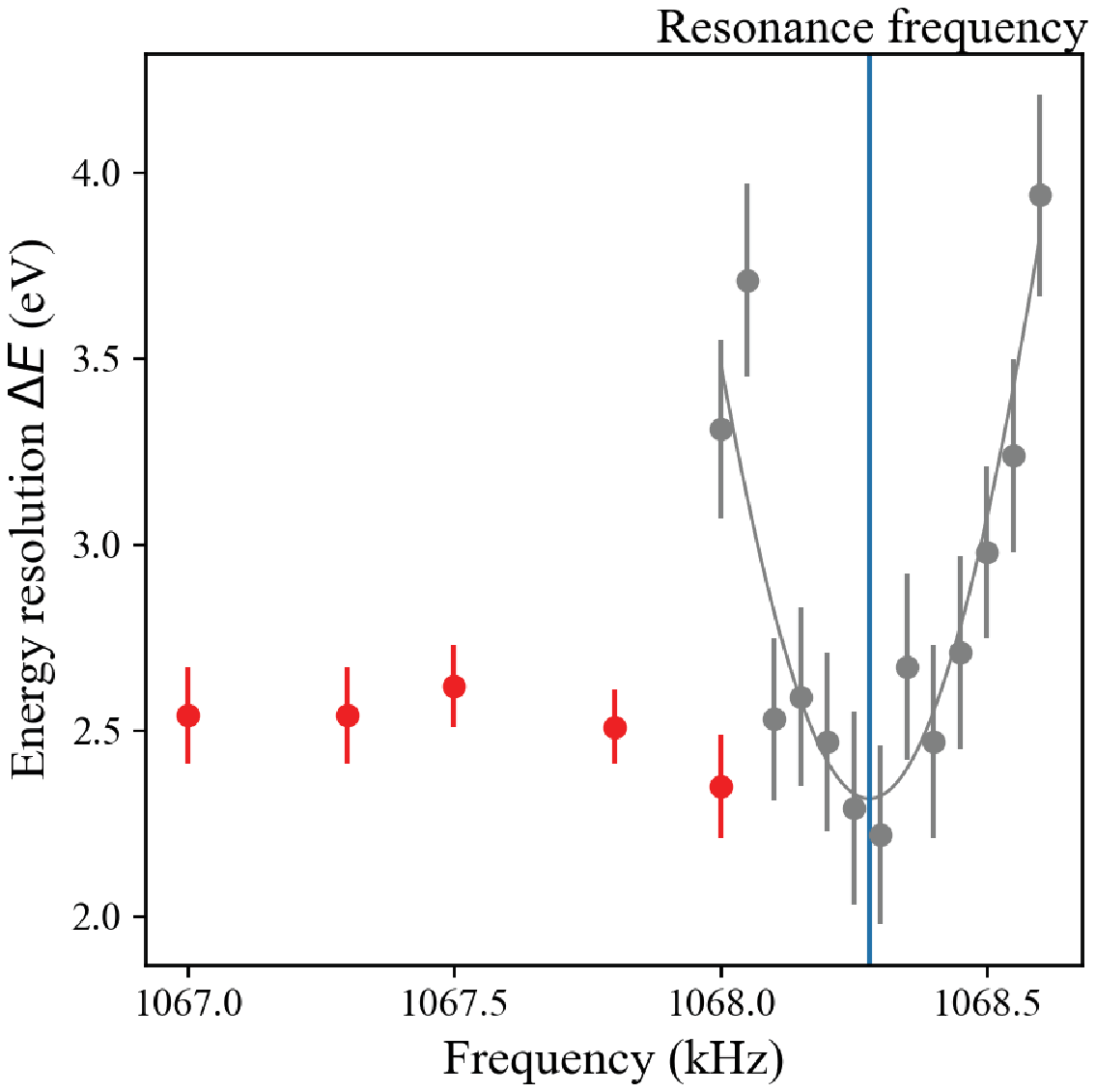}
\includegraphics[width=0.485\linewidth, keepaspectratio]{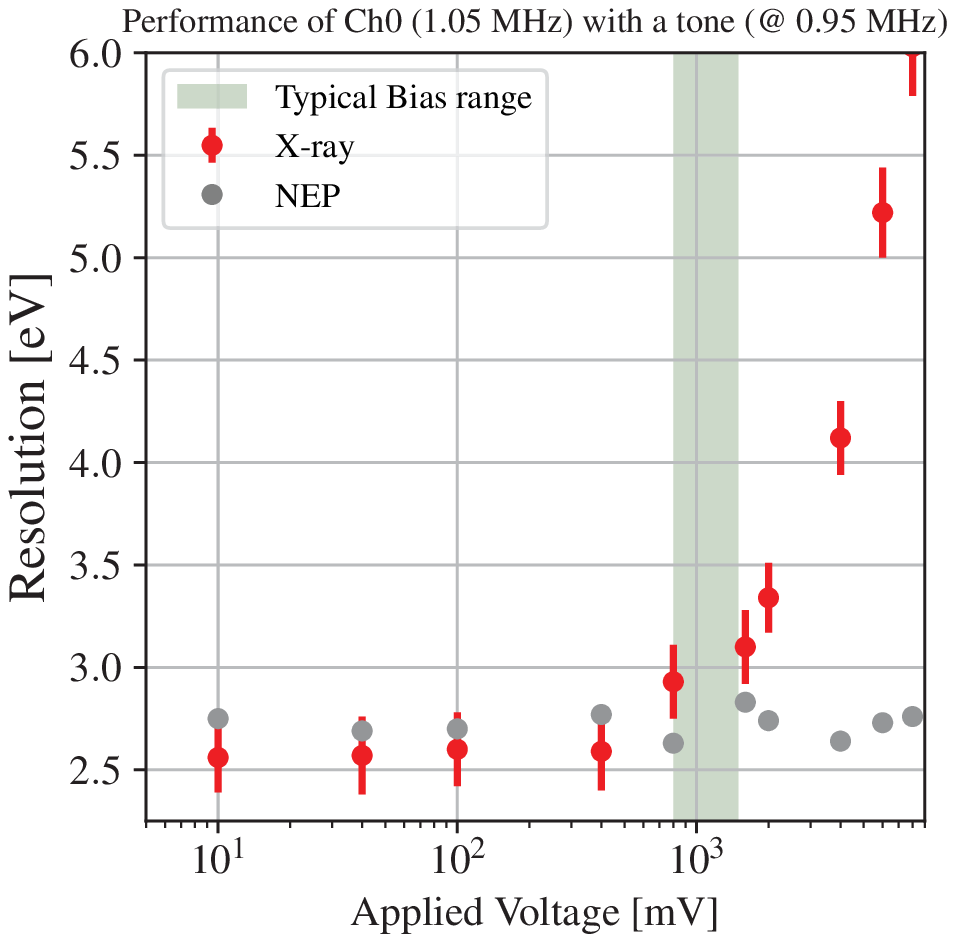}
\caption{ \label{fig:FSA}
(Left) Single pixel X-ray performance with (red) and without (gray) FSA. The resonant frequency is indicated by the blue vertical line. 
(Right) Single pixel X-ray performance with additional AC bias tone at 100 kHz lower frequency. Gray and red points indicate integrated NEP and X-ray resolution, respectively. Green shed region indicate typical bias voltage for $R\sim 0.2 Rn$.
}
\vspace{-0.5cm}
\end{center}
\end{figure}

To confirm this effect, we perform an experiment: single pixel X-ray measurement with additional tone at 100 kHz separated frequency and different bias voltages. 
\blue{The right panel in Fig.\ref{fig:FSA}} shows the results. As expected, the NEP resolutions are  constant across different bias voltage. On the other hand, X-ray resolutions show a strong dependence on the applied voltages.  When the applied voltage is small, there is no effect on the performance. However, once the voltage is getting close to the typical TES voltages (green shaded area), the performances degrade significantly. This means TES has still too large electrical bandwidth to be independent from neighboring  bias voltages. 
\blue{This means that TES has still dependence on the neihboring voltages because of too large electrical bandwidth.}

This effect can be solved in three ways: 1.) use a larger frequency separation, 2.) apply  a phase window \cite{2003ITAS...13..638V} and 3.) use a narrower electrical bandwidth detector.  From X-IFU instrument point of view, option 1 and 2 will not be the final solution. Therefore, option 3 will be the solution. The electrical bandwidth can be suppressed by making TES critically damped. However, as shown in Table\ref{tab:device}, TESs are working close to critical damping at 14 pixel MUX experiment. To narrow down the electrical bandwidth, 
\blue{
we need different devices, which have much slower detector response time, allowing us to use larger inductance. 
}
Here we note that detector designs of GSFC-A6 and A7 are a factor 2--3 faster than X-IFU requirements. 
Currently, new devices which will satisfy the X-IFU requirements and further larger setups are under development. Once they are ready we are expecting that FDM demonstration will have another jump in terms of the number of multiplexed pixels. 
\vspace{-.75cm}

\section{Summary}\vspace{-0.25cm}
We are developing a Frequency domain multiplexing technology for the X-IFU on board the Athena mission. By using the state-of-art TESs, cryogenic and room temperature electronics, we confirmed that the TES performance under AC bias is compatible with DC bias case ($\Delta E\sim 1.8$ eV). We also demonstrated 9 (200 kHz) and 14 (100 kHz) pixel MUX readouts with 2.6 and 3.3 eV summed performances, respectively. The observed degradation from 200 kHz to 100 kHz frequency separation can be explained by 1.) impact of the intermodulation line noise (sect.\ref{sec:line}) and 2.) Influences from neighboring AC bias voltage (sect.\ref{sec:ACB}).  Future X-IFU device, which will be a factor 2--3 slower than current one, will improve the situation in both intermodulation line noises and interference from neighboring bias voltage.

\begin{acknowledgements}
The authors would like to thank Martijn Schoemans, 
 Dick Boersma, Marcel van Litsenburg, Patrick van Winden and Bert-Joost van Leeuwen
for their precious/continues help. The authors also thank
an anonymous referee for constructive comments. SRON is
supported financially by NWO, the Netherlands Organization
for Scientific Research.

\end{acknowledgements}


\begin{thebibliography}{99}

\bibitem[Barret et al.(2019)]{2019arXiv191204615B} Barret, D., Decourchelle, A., Fabian, A., et al.\ 2019, arXiv e-prints, arXiv:1912.04615


\bibitem[Barret et al.(2018)]{XIFU18} Barret, D., Lam Trong, T., den Herder, J.-W., et al.\ 2018, SPIE, 106991G

\bibitem[Rau et al.(2013)]{WFI13} Rau, A., Meidinger, N., Nandra, K., et al.\ 2013, arXiv e-prints, arXiv:1308.6785

\bibitem[Gottardi et al. (2019)]{luciano_LTD18_AC} L. Gottardi et al. these proceedings

\bibitem[Akamatsu et al.(2014)]{akamatsu14} Akamatsu, H., Gottardi, L., Adams, J., et al.\ 2014, Journal of Low Temperature Physics, 176, 591

\bibitem[Akamatsu et al.(2016)]{akamatsu16} Akamatsu, H., Gottardi, L., de Vries, C.~P., et al.\ 2016, Journal of Low Temperature Physics, 184, 436

\bibitem[]{LCcooler} https://leiden-cryogenics.com/


\bibitem[Bruijn et al.(2018)]{LCfilter18} Bruijn, M.~P., van der Linden, A.~J., Ferrari, L., et al.\ 2018, Journal of Low Temperature Physics, 193, 661

\bibitem[van der Kuur et al.(2016)]{2016SPIE.9905E..5RV} van der Kuur, J., Gottardi, L.~G., Akamatsu, H., et al.\ 2016, SPIE, 99055R


\bibitem[Gottardi et al.(2019)]{luciano_vib19} Gottardi, L., van Weers, H., Dercksen, J., et al.\ 2019, Review of Scientific Instruments, 90, 055107

\bibitem[den Hartog et al.(2009)]{BBFB} den Hartog, R., Boersma, D., Bruijn, M., et al.\ 2009, American Institute of Physics Conference Series, 261

\bibitem[Kaastra(2017)]{cstat} Kaastra, J.~S.\ 2017, A\&A, 605, A51

\bibitem[Smith et al.(2016)]{smith16} Smith, S.~J., Adams, J.~S., Bandler, S.~R., et al.\ 2016, SPIE, 99052H

\bibitem[Miniussi et al.(2018)]{2018JLTP..193..337M} Miniussi, A.~R., Adams, J.~S., Bandler, S.~R., et al.\ 2018, Journal of Low Temperature Physics, 193, 337

\bibitem[Sakai et al.(2018)]{2018JLTP..193..356S} Sakai, K., Adams, J.~S., Bandler, S.~R., et al.\ 2018, Journal of Low Temperature Physics, 193, 356

\bibitem[Gottardi et al.(2018)]{2018JLTP..193..209G} Gottardi, L., Smith, S.~J., Kozorezov, A., et al.\ 2018, Journal of Low Temperature Physics, 193, 209

\bibitem[van der Kuur et al.(2003)]{2003ITAS...13..638V} van der Kuur, J., de Korte, P.~A.~J., Hoevers, H.~F.~C., et al.\ 2003, IEEE Transactions on Applied Superconductivity, 13, 638

\bibitem[van der Kuur et al.(2018)]{2018JLTP..193..626V} van der Kuur, J., Gottardi, L., Akamatsu, H., et al.\ 2018, Journal of Low Temperature Physics, 193, 626


\end{thebibliography}
\end{document}